# Can Frequency Diverse Array Prevent Wireless Eavesdropping in Range Domain?

Yuan Ding, *Member*, *IEEE*, and Adam Narbudowicz, *Member, IEEE*

*Abstract*— In this paper, the concept and recent development of exploiting frequency diverse array (FDA) and its variants for the physical-layer wireless security have been revisited and carefully examined. Following rigorous analytical derivation and illustrative simulations, the authors argue that the investigations performed in some recent works did not reveal one critical issue facing the real-world applications, and system models established and used before were based on an unrealistic assumption, i.e. that the legitimate and eavesdropping users at different ranges sample the signal waveforms at the same time instant. This misunderstanding results in conclusions that are misleading. The authors aim to take the first step to divert research efforts and rectify the previous problematic analyses. The authors prove that the FDA cannot secure a free-space wireless transmission in range domain, because the previously claimed 'secure reception region' propagates in range domain as time elapses.

## I. INTRODUCTION

Directional Modulation (DM) is a type of keyless physical-layer wireless security technique, which, in its original form, is able to transmit desired signal waveforms with information modulated only along pre-selected directions, while distorting the waveforms along all other spatial directions in free space [1]–[4]. In this fashion the information transmitted wirelessly in free space can only be reliably recovered by legitimate receivers positioned along those selected directions, enhancing security directly at physical layer. The technique is very attractive for security applications of modern radio systems, as it does not require mathematically generated cryptographic keys and supports a very simple receiver's architecture. However, the biggest issue for real-life applications is its inability to provide security in range-domain when line-of-sight (LoS) communication links are concerned, i.e. any eavesdropper located at the same direction as the legitimate receiver is able to intercept the information that is supposed only for the legitimate receiver.

Recent theoretical works in [5]–[15] attempted to solve this problem. The proposed solutions combine DM with a Frequency Diverse Array (FDA) – a technique used in radar systems to illuminate target at a given range with a multi-frequency signal of short duration. If successful, the combination of FDA-DM would have allowed unprecedented levels of wireless security, offering to securely transmit information to almost any wireless device without the need for traditional cryptographic encryption algorithms, avoiding problem of

Manuscript received Sep. 26, 2019.
Y. Ding is with the Institute of Sensors, Signals and Systems (ISSS), Heriot-Watt University, Edinburgh, United Kingdom, EH14 4AS (e-mail: yuan.ding@hw.ac.uk, tel: +44(0)131 451 4155)
A. Narbudowicz is with Department of Telecommunications and Teleinformatics, Wroclaw University of Science and Technology, Wroclaw, Poland, 50-370 (e-mail: adam.narbudowicz@pwr.edu.pl)

key distribution.

Despite promising theoretical results, to the best of the authors' knowledge, currently there is no experimental validation of the FDA-DM that demonstrates its alleged security in range domain. This mismatch between theoretical and experimental work decelerates further development of DM, as the combination of FDA with DM so far did not deliver on its promise of security without encryption. However, a brief observation was made in [16], [17] relating to the potential problem of time-invariance of FDA systems in the context of radiation-pattern optimization.

This paper provides the first in-depth analysis into the mechanisms of the FDA-DM security in range domain. It significantly extends the analyses provided in [5]–[15] in order to include a more generic approach with time as the third missing variable. Proposed analysis demonstrates that the so obtained security in range-domain cannot be time-invariant and that the "secure region" propagates with time. The provided results are generic for any FDA-based security technique, regardless of the antenna structure and secrecy metric used.

## II. FREQUENCY DIVERSE ARRAY

The concept of FDA was first introduced in [18], [19]. It employs array elements that radiate electromagnetic waves with slightly different frequencies, where the frequency differences are assumed to be many orders smaller than the reference carrier frequency. Fig. 1 illustrates a one-dimensional (1D) uniformly spaced *N*-element FDA with a linear carrier frequency increment $\Delta f$ applied across the array. The first array element is taken as the reference with its excitation at the carrier frequency $f_0$. Here $\Delta f << f_0$.

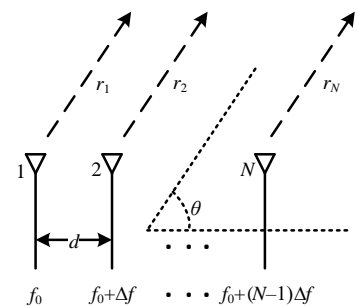

Fig. 1. Illustration of a 1D FDA with uniform frequency increment $\Delta f$.

It was presented in [18] and [19] that the '*beamforming patterns*' are angle-range dependent. Two aspects need to be emphasized;
1) The term '*beamforming patterns*' used in [18] and [19] is not the same concept as the '*far-field radiation patterns*' commonly used in the literature on antenna engineering, since the latter by definition refers to single frequency [20]. The '*beamforming patterns*' can be instead interpreted as 'normalized electric (or magnetic) field distribution in the far-field';
2) The '*beamforming patterns*' are also a function of time. Since in [18] and [19] the radiated waveforms were designed to be very short

pulses for Radar applications, the time instant *t* associated with the detected electric field at a certain distance *R* is thus uniquely determined, i.e. $t = R/c$ where *c* is the speed of light. In other words, the '*beamforming patterns*' shown in [18] and [19] are not snap-shots at a certain time instant, but a series of snap-shots presented as a function of single parameter that combines time and range as $t = R/c$. This angle-range dependent '*beamforming patterns*' of the pulsed FDA have been extensively investigated since then, resulting in a number of promising pulsed FDA Radar systems, seen in [21]–[23] and references therein. However, contrary to Radar applications, in order to establish a wireless communication link (the scope of the FDA systems studied in this paper), a continuous electromagnetic wave needs to be transmitted to carry information. Thus, the detected electric (or magnetic) field at each angle-range coordinate varies with time, i.e. '*beamforming patterns*' are functions of angle, range and time.

In the meantime, another research effort on Directional Modulation (DM) was made in the antenna and propagation community. Most early DM works could only securely transfer narrow band signals. Therefore, the authors in [24] made the first effort in combining FDA with DM so that an OFDM modulated DM system was constructed. Here, the FDA transmits signal waveforms continuously in time, different to pulsed signals for FDA Radar applications. No range-domain security was claimed in [24].

Inspired by the FDA range-angle dependent beamforming patterns, and its first introduction to the DM systems, many recent research efforts have been focused on using FDA concept to secure free space DM systems in range domain, e.g. [5]–[7], resulting so-called FDA-DM systems. However, an important factor was overlooked, since the FDA range-angle dependent beamforming patterns are also functions of time. This indicates that the secure reception regions (normally defined as the locations where the received bit error rates (BERs) are below a specified threshold) propagate in range as time elapses. This is analyzed in more detail in the subsequent section.

## III. DM AND FDA IN THEIR GENERAL FORMS

In this section, we present mathematical modelling of both DM and FDA in their general forms, from which the previously reported FDA-DM fusion systems can be derived in Section III, leaving their discussions revealing the flaws in Section IV.

### A. DM

For a 1D *N*-element transmit array, e.g. the one shown in Fig. 1 with $\Delta f$ being set to zero, the received far-field electric (or magnetic) field **F** can be expressed as

$$\begin{aligned} \mathbf{F} &= \sum_{n=1}^{N} \mathbf{G}_n \cdot \exp\left[j2\pi f_0\left(t - \frac{r_1 - (n-1)d\cos\theta}{c}\right)\right] \\ &= \sum_{n=1}^{N} \mathbf{G}_n \cdot \underbrace{\exp\left(\frac{j2\pi f_0(n-1)d\cos\theta}{c}\right)}_{H_n(\theta)} \cdot \exp\left[j2\pi f_0\left(t - \frac{r_1}{c}\right)\right] \\ &= \vec{\mathbf{G}}\vec{\mathbf{H}}^T(\theta) \cdot \exp\left[j2\pi f_0\left(t - \frac{r_1}{c}\right)\right] \end{aligned} \quad , \quad (1)$$

where $\vec{\mathbf{H}} = [\mathbf{H}_1, \mathbf{H}_2, \ldots, \mathbf{H}_N]$ is the channel vector, and '$[\cdot]^T$' refers to vector transpose operator. *d* is the uniform spacing between two consecutive array elements. $r_1$ denotes the displacement between the first antenna (as the reference) and the far-field observation point. $\theta$ is the spatial direction with respect to the array, ranging from 0 to $\pi$ defined in Fig. 1. In order to achieve DM functionality [2], in general form the array excitation vector $\vec{G}$ is designed to be

$$\vec{G} = p\mathbf{D}\vec{\mathbf{H}}^*(\theta_0) + q\vec{W}. \quad (2)$$

Here **D** is a complex number, representing a symbol (i.e. information modulated in IQ space) intended for transmission. Vector $\vec{W}$ is power normalized, i.e. $\vec{W}\vec{W}^\dagger = 1$, and it lies in the null space of channel vector $\vec{\mathbf{H}}(\theta_0)$, i.e. $\vec{W}\vec{\mathbf{H}}^T(\theta_0) = 0$. '$[\cdot]^*$' and '$[\cdot]^\dagger$' denote conjugation and vector Hermitian transpose, respectively. $\theta_0$ is the desired secure communication direction. *p* and *q* determine the power allocation between useful information **D** and orthogonal artificial noise $\vec{W}$.

When inserting (2) into (1), we get

$$\begin{cases} \mathbf{F} = pN\mathbf{D} \cdot \exp\left[j2\pi f_0\left(t - \frac{r_1}{c}\right)\right] & \text{when } \theta = \theta_0 \\ \mathbf{F} = \left[p\mathbf{D}\vec{\mathbf{H}}^*(\theta_0) \cdot \vec{\mathbf{H}}^T(\theta) + q\vec{W} \cdot \vec{\mathbf{H}}^T(\theta)\right] \\ \quad \cdot \exp\left[j2\pi f_0\left(t - \frac{r_1}{c}\right)\right] & \text{when } \theta \neq \theta_0 \end{cases}. \quad (3)$$

From (3), we can see that the information **D** (or associated modulation waveforms) are transferred to the legitimate user along $\theta = \theta_0$, while for other directions the information **D** are contaminated with the randomly updated artificial noise $\vec{W}$, greatly reducing probability of interception.

### B. FDA

In this subsection, we mathematically describe how an FDA, shown in Fig. 1, operates. The radio frequency (RF) carrier frequency applied at each antenna element is

$$f_n = f_0 + (n-1)\cdot\Delta f, \qquad n = 1, 2, \ldots, N \quad (4)$$

For this uniformly spaced 1D FDA array, the received (pathloss being normalized out) far-field electric (or magnetic) field **B** along a spatial direction $\theta$ in free space can be written as

$$\mathbf{B} = \sum_{n=1}^{N} \mathbf{A}_n \exp\left[\underbrace{j2\pi f_n\left(t - \frac{r_1 - (n-1)d\cos\theta}{c}\right)}_{\phi_n}\right], \quad (5)$$

where $\mathbf{A}_n = \mathbf{D}v_n\exp(j\varphi_n)$ is the excitation of the $n^{th}$ array element, with scalar valued $v_n$ of amplitude and $\varphi_n$ of phase in its initial state. Without loss of generality, $v_n\exp(j\varphi_n)$ is set to be unity for each *n*.

The phase term $\phi_n$, seen in (5) can be further expressed as

$$\begin{aligned} \phi_n &= 2\pi\left[f_0 + (n-1)\Delta f\right]\left(t - \frac{r_1}{c} + \frac{(n-1)d\cos\theta}{c}\right) \\ &= 2\pi f_0\left(t - \frac{r_1}{c}\right) + \left[\frac{2\pi f_0(n-1)d\cos\theta}{c} + 2\pi(n-1)\Delta ft \right. \\ &\quad \left. -\frac{2\pi(n-1)\Delta f r_1}{c} + \frac{2\pi(n-1)^2\Delta fd\cos\theta}{c}\right]. \end{aligned} \quad (6)$$

The last term in the bracket in (6) is extremely tiny for practical FDA configurations. For example, when $\Delta f = 10$ kHz, $f_0 = 3$ GHz, $d =$ $c/(2f_0)$, and $N = 10$, this last phase term is less than 0.05°. This term is thus omitted hereafter. The phase difference with respect to the signal radiated by the first antenna is

$$\Delta_n = \phi_n - \phi_1$$
$$= \frac{2\pi f_0 (n-1) d \cos\theta}{c} - \frac{2\pi (n-1) \Delta f r_1}{c} + 2\pi (n-1) \Delta f \cdot t. \quad (7)$$

The beamforming pattern $\boldsymbol{B}$ in (5) reaches its peaks when the phases of every summation terms are aligned. It requires $\Delta_n$ to be $2k_n\pi$ for each $n$. $k_n$ can be any arbitrary integers. Equivalently,

$$\frac{f_0 d \cos\theta}{c} - \frac{\Delta f r_1}{c} + \Delta f t = z. \quad z = 0, \pm 1, \pm 2, ... \quad (8)$$

This shows how the beamforming pattern peaks change with angle $\theta$, range $r_1$, and time $t$.

When (8) is satisfied,

$$\boldsymbol{B} = N\boldsymbol{D} \cdot \exp\left[j2\pi f_0 \left(t - \frac{r_1}{c}\right)\right], \quad (9)$$

which means the information $\boldsymbol{D}$ is delivered to the coordinate $(\theta, r_1)$ at the time instant $t$, with a beamforming gain of $20\times\log_{10}(N)$ in dB. Since $z$ can be any integer, there are infinite solutions to (8), meaning no wireless security can be achieved with FDA alone.

## IV. Previously Reported FDA-DM Systems

When realizing FDAs can generate angle-range dependent beamforming patterns, plenty of efforts have been made to incorporate FDA concept into DM transmitters [6]–[11], claiming that the information $\boldsymbol{D}$ can be securely delivered to a pre-specified angle-range coordinate, saying $(\theta_0, R)$. The general form of the resulting FDA-DM systems reported in previous works is formulated in this section.

**Note:** In Section IV, the authors will argue that the resulting FDA-DM systems, however, **CANNOT** secure information in range domain as what the reported works have claimed. This is because an important fact that FDA beamforming patterns at each spatial location are time-dependent was overlooked.

Combining FDA and DM, namely applying baseband DM excitation vector $\vec{G}$ in (2) onto the frequency shifted RF carriers in (5), the electric (or magnetic) field in any far-field location $(\theta, r_1)$ becomes

$$\boldsymbol{E} = \sum_{n=1}^{N} G_n \exp\left[j2\pi f_n \left(t - \frac{r_1 - (n-1)d\cos\theta}{c}\right)\right]. \quad (10)$$

Here $G_n$ is the $n^{th}$ entry of the vector $\vec{G}$. Replacing $G_n$ in (2) into (10), we get

$$\boldsymbol{E} = \sum_{n=1}^{N}\left\{p\boldsymbol{D} \cdot \exp\left[\frac{j2\pi f_0(n-1)d(\cos\theta - \cos\theta_0)}{c}\right]\right.$$
$$\left. \cdot \exp\left[j2\pi(n-1)\Delta f\left(t - \frac{r_1}{c}\right)\right] \cdot \exp\left[j2\pi f_0\left(t - \frac{r_1}{c}\right)\right]\right\}$$
$$+ \sum_{n=1}^{N}\left\{q\boldsymbol{W}_n \cdot \exp\left[\frac{j2\pi f_0(n-1)d\cos\theta}{c}\right]\right.$$
$$\left. \cdot \exp\left[j2\pi(n-1)\Delta f\left(t - \frac{r_1}{c}\right)\right] \cdot \exp\left[j2\pi f_0\left(t - \frac{r_1}{c}\right)\right]\right\}, \quad (11)$$

where $\boldsymbol{W}_n$ is the $n^{th}$ entry of the vector $\vec{W}$. The same tiny phase as the last term in (6) is safely ignored here.

The previously reported FDA-DM works [5]–[11] claimed that for a desired receiver's located at $(\theta_0, R)$ the information $\boldsymbol{D}$ can be uniquely conveyed to the desired receiver at $(\theta_0, R)$ only. This is under assumption that the receiver samples signals at a reference time $t = 0$, when $\Delta f$ of the transmitted signal is designed to be $c/R$. In these conditions, (11) becomes

$$\boldsymbol{E}(\theta = \theta_0, r_1 = R, t = 0) = pN\boldsymbol{D} \cdot \exp\left(-j2\pi f_0 \frac{R}{c}\right). \quad (12)$$

For locations other than $(\theta_0, R)$, the second summation in (11) is non-zero at $t = 0$, acting as orthogonal artificial noise in both angle and range domains.

A simulation example of (11) is illustrated in Fig. 2 with FDA-DM system parameters listed in Table I. From Fig. 2, it can be seen that the QPSK modulated waveforms are only preserved in a pre-identified location ($\theta_0 = 40°$, $R = 30$ km), achieving so-called secure wireless transmission in both angle and range domains. Two special cases are: *a*) when $\Delta f = 0$, the second summation term in (11) is not a function of range (excluding the identical time harmonic term at $f_0$ for every $n$), indicating no artificial noise is injected in range domain; *b*) when $q = 0$ (irrespective of the choice of $\Delta f$), the second summation term in (11) is zero, indicating no artificial noise is injected in both angle and range domains. Readers can perform the simulations to verify the special cases if interested. The results are omitted here for brevity. The secure reception region can be further shrunk by increasing the number of antenna elements $N$ and/or by allocating more power to artificial noise (i.e. increasing the ratio $q/p$), which are common strategies used in DM systems [2].

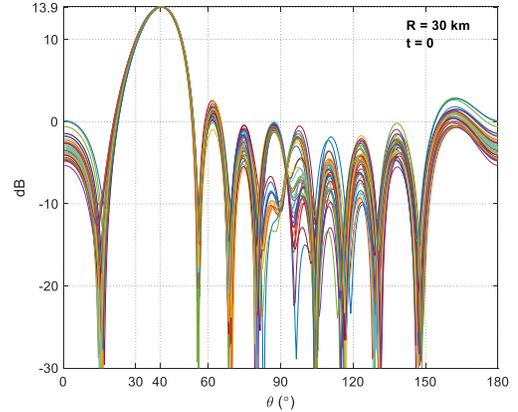

(a)

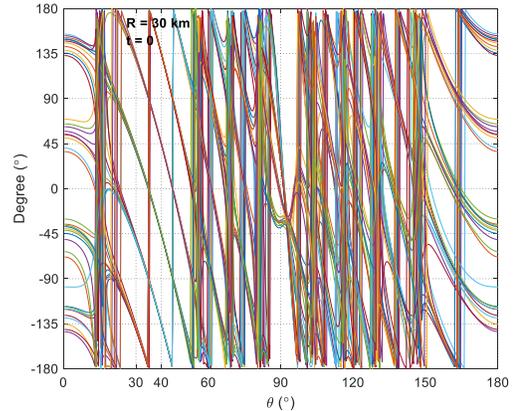

(b)

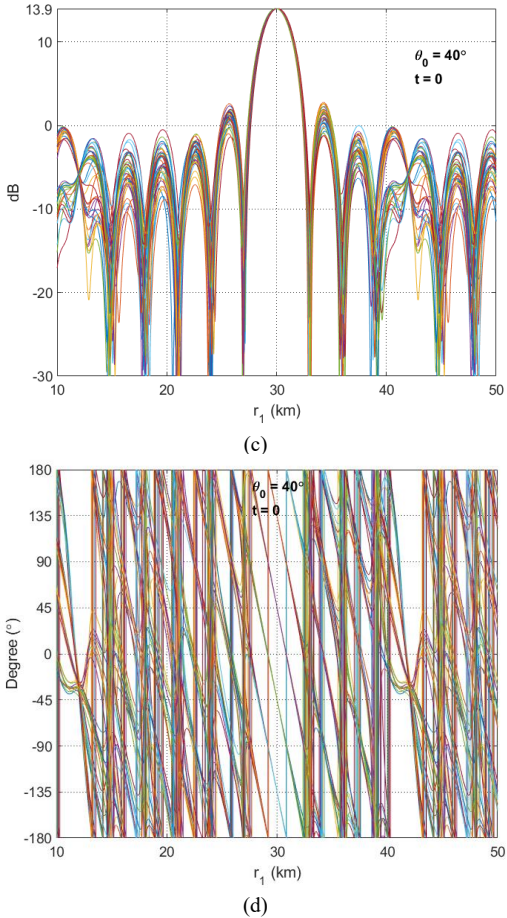

Fig. 2. Simulation example of previously reported FDA-DM system in its general form in (11). System parameters are listed in Table I. (a) Magnitudes and (b) phases of electric (or magnetic) fields in angle domain when $r_1 = R = 30$ km; (c) Magnitudes and (d) phases of electric (or magnetic) fields in range domain when $\theta = \theta_0 = 40°$ (pathloss is removed for illustration purpose).

TABLE I
SYSTEM PARAMETERS OF A PREVIOUSLY REPORTED FDA-DM EXAMPLE IN ITS GENERAL FORM IN (11)

| Parameter | Value |
| --- | --- |
| $f_0$ | 3 GHz |
| $R$ | 30 km |
| $\theta_0$ | 40° |
| $\Delta f$ | $c/R$ = 10 kHz |
| $N$ | 10 |
| $d$ | $0.5 \times c/f_0$ = 5 cm |
| $p$ | 0.5 |
| $q$ | 0.5 |
| $t$ | 0 |
| $D$ | Randomly generated QPSK symbols $\{\exp(j\pi/4); \exp(j3\pi/4); \exp(-j\pi/4); \exp(-j3\pi/4)\}$ |
| Number of simulated symbols | 40 |

The above common proposition claimed in previous FDA-DM works will be rebutted in the following section.

## V. SECURE RECEPTION REGIONS 'PROPAGATE' IN RANGE AS TIME ELAPSES IN FDA-DM SYSTEMS

In this section, the authors argue that the constructed FDA-DM systems, formulated in (11), **CANNOT** secure wireless transmissions in range domain. The misinterpretation of (11) was rooted in the treatment of time $t$. Like the example shown in Fig. 2, the previous FDA-DM works use far-field patterns in 2D angle-range domain **at a selected time reference** when the legitimate receiver samples detected signals, to claim the secure transmission in range domain. Thus, the patterns, such as those plotted in Fig. 2, are the field distributions at that chosen time instant.

Two problems are associated with this time treatment;
1) At the legitimate receiver end, in order to perform demodulation, the entire modulation symbol with a symbol period $T$ is frequency down-converted first, before baseband sampling. Within this $T$, the term $\exp[j2\pi(n-1)\Delta f(t-r_1/c)]$ in the second summation in (11) is not identical for each $n$, indicating the artificial noise cannot be perfectly cancelled out at the selected location ($\theta_0$, $R$). The amount of the remaining artificial noise for the legitimate receiver is determined by the system parameters, in particular $N$, $\Delta f$, and $T$. This aspect has never been studied, while the second problem discussed below is more critical, which invalidates the previously reported FDA-DM systems;

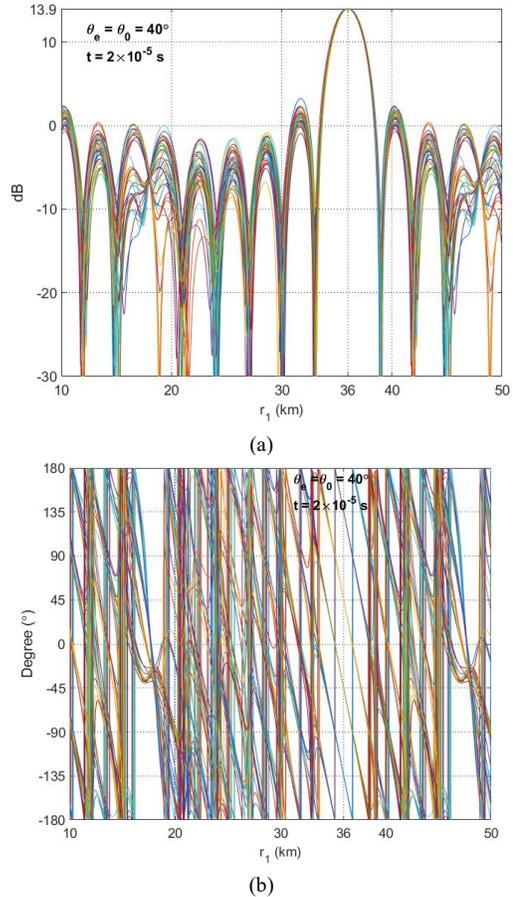

Fig. 3. Simulation example of previously reported FDA-DM system in its general form in (11). System parameters are listed in Table I (except time $t$). (a) Magnitudes and (b) phases of electric (or magnetic) fields in range domain when $\theta = \theta_0 = 40°$ and $t = 2\times10^{-5}$ s (pathloss is removed for illustration purpose).

2) The early FDA-DM works used the beamforming patterns at the selected time instant to calculate secure reception regions, claiming range-domain wireless security. **However, the eavesdroppers do not necessarily sample the signals at the same time instant as the desired receiver does.** Taking the same FDA-DM example with the settings in Table I (except time $t$), the simulated range-domain far-field patterns at the time instant $t = (R_e − R)/c = $ (36 km – 30 km)$/c = 2\times10^{-5}$ s are illustrated in Fig. 3. It is assumed that an eavesdropper receiver is positioned at ($\theta_e = \theta_0 = 40°$, $R_e = 36$ km). From Fig. 3, it can be clearly seen that the secure reception region (for this example it is the location where the well-formatted QPSK IQ constellations/waveforms are preserved) 'propagates' at the speed of light, as time elapses. Similarly, when $\theta_e = \theta_0$ and $R_e < R$, the well-preserved signal waveforms pass the eavesdropper at the time instant $(R − R_e)/c$ ahead the time reference. To conclude, **NO secure transmissions in range domain can be achieved by the previously reported FDA-DM systems**.

**Another intuitive explanation** of why the FDA-DM systems cannot provide range-domain security is presented below:

Assuming a legitimate receiver positioned at ($\theta_0$, $R$) in free space detects electromagnetic waves which correspond to desired modulation symbols, these electromagnetic waves, spatially combined by each electromagnetic wave radiated from each transmit antenna, propagate at the speed of light along $\theta_0$, irrespective of their frequencies. Therefore, the same signal waveforms (subject to magnitude scaling) detected by the legitimate receiver reach every points along $\theta_0$ at different time instants when the far-field condition is met. From this observation, it can be concluded that any FDA-DM arrangements, including their variants **CANNOT** provide secure wireless transmission in range domain in free space.

Next, we briefly list the issues in some recent FDA-DM literatures;
- [5]–[11]:

A time dependent phase term (i.e. $\alpha_n$ shown in (7) in this paper) was missing in {(5) in [5]; (4) in [6]; (2) in [7]; (3) in [8]; (6) in [9]; (4) in [10]; (2) in [11]} and all the analyses thereafter. This indicates that an assumption $t = 0$ was made for every receiver in the field. In other words, the authors in these works enforce legitimate and eavesdropping receivers sampling signals at the same instant, leading to erroneous conclusions;
- [12]:

The authors used the same time instant to sample the received signals at both legitimate and eavesdropping receivers, see (11) and (12) in [12];
- [13]:

From (10) in [13], the author claimed that the radiation energy is focused at ($\theta^{in}$, $R$). Mathematically, this can only be obtained when $t = 0$. In fact, when $t$ varies, it can be seen that this focusing point propagates at the speed of light in range domain;
- [14], [15]:

In {(9) in [14]; (4) in [15]}, the time '$t$' in the denominator is the time reference the authors selected (when the legitimate receiver samples signals), while the time '$t$' in the numerator should be the time instant when each receiver samples their detected signals. These two time '$t$' are not necessarily identical. In fact, when different '$t$' in numerator is chosen, it can be observed that the spatial focusing region propagates;
- [25]:

In fact, the Fig. 2 in [25] and the associated discussions clearly shown that the secure reception region propagates at the speed of light in range domain. However, the authors claimed that if the array excitation vector changes accordingly, the secure region does not propagate. This statement is erroneous, as the continuously altered array radiation at the transmitter end cannot instantly propagates through space. When the propagation delay is considered, it can be observed that the secure reception region propagates no matter the excitation vector changes or not.

Based on the above analyses, the conclusions reached in some previous FDA works [5]–[15], [25] are unreliable.

## VI. CONCLUSION

The realization of DM scheme that allows security with respect to both direction and range domains is a significant research problem, with potentially high impact if solved. However, the paper demonstrated that such security cannot be obtained by combining DM with FDA when the time variable is incorporated in the investigated model. It has been demonstrated that the 'secure area' will propagate in range – similarly to any other electromagnetic signal – and consequently any eavesdropper located along the pre-defined direction is able to easily intercept the signal within limited time.

REFERENCES

[1] M. P. Daly and J. T. Bernhard, "Directional modulation technique for phased arrays," *IEEE Trans. Antennas Propag.*, vol. 57, no. 9, pp. 2633–2640, Sep. 2009.
[2] Y. Ding and V. Fusco, "A vector approach for the analysis and synthesis of directional modulation transmitters," *IEEE Trans. Antennas Propag.*, vol. 62, no. 1, pp. 361–370, Jan. 2014.
[3] Y. Ding and V. Fusco, "A review of directional modulation technology," *Int. J. Microw. Wireless Technol.*, vol. 8, no. 7, pp. 981–993, 2016.
[4] Y. Ding and V. Fusco, "A synthesis-free directional modulation transmitter using retrodirective array," *IEEE J. Sel. Topics Signal Process.*, vol. 11, no. 2, pp. 428–441, Mar. 2017
[5] S. Y. Nusenu and A. Basit, "Frequency diverse array antennas: from their origin to their application in wireless communication systems," *J. Computer Netw. Commun.*, vol. 2018, Article ID 5815678, 12 pages, 2018.
[6] S. Y. Nusenu, W. Wang, and S. Ji, "Secure directional modulation using frequency diverse array antenna," in *IEEE Radar Conf.*, Seattle, WA, 2017, pp. 0378–0382.
[7] J. Xiong, S. Y. Nusenu, and W. Wang, "Directional modulation using frequency diverse array for secure communications," *Wireless Pers. Commun.*, vol. 95, no. 3, pp. 2679–2689, Aug. 2017.
[8] J. Xiong, W. Wang, H. Shao, and H. Chen, "Frequency diverse array transmit beampattern optimization with genetic algorithm," *IEEE Antennas Wireless Propag. Lett.*, vol. 16, pp. 469–472, Aug. 2017.
[9] J. Hu, S. Yan, F. Shu, J. Wang, J. Li, and Y. Zhang, "Artificial-noise-aided secure transmission with directional modulation based on random frequency diverse arrays," *IEEE Access*, vol. 5, pp. 1658–1667, 2017.
[10] S. Y. Nusenu, H. Chen, W. Wang, and S. Ji, and O. Opuni-Boachie, "Frequency diverse array using Butler Matrix for secure wireless communications," *Progress In Electromagn. Research M*, vol. 63, 207–215, 2018.
[11] S. Y. Nusenu and W. Wang, "Range-dependent spatial modulation using

frequency diverse array for OFDM wireless communications," *IEEE Trans. Veh. Technol.*, vol. 67, no. 11, pp. 10886–10895, Nov. 2018.

[12] F. Shu, X. Wu, J. Hu, J. Li, R. Chen, and J. Wang, "Secure and precise wireless transmission for random-subcarrier-selection-based directional modulation transmit antenna array," *IEEE J. Sel. Areas Commun.*, vol. 36, no. 4, pp. 890–904, Apr. 2018.

[13] W. Wang, "Retrodirective frequency diverse array focusing for wireless information and power transfer," *IEEE J. Sel. Areas Commun.*, vol. 37, no. 1, pp. 61–73, Jan. 2019.

[14] A. Yao, W. Wu, and D. Fang, "Solutions of time-invariant spatial focusing for multi-targets using time modulated frequency diverse antenna arrays," *IEEE Trans. Antennas Propag.*, vol. 65, no. 2, pp. 552–566, Feb. 2017.

[15] Y. Yang, H. Wang, H. Wang, S. Gu, D. Xu, and S. Quan, "Optimization of sparse frequency diverse array with time-invariant spatial-focusing beampattern," *IEEE Antennas Wireless Propag. Lett.*, vol. 17, no. 2, pp. 351–354, Feb. 2018.

[16] M. Fartookzadeh, "Comments on 'Optimization of sparse frequency diverse array with time-invariant spatial-focusing beampattern'," *IEEE Antennas Wireless Propag. Lett.*, vol. 17, no. 12, pp. 2521–2521, Dec. 2018.

[17] Y.-Q. Yang, H. Wang, H.-Q. Wang, S.-Q. Gu, D.-L. Xu, and S.-L. Quan, "Reply to "Comments on 'Optimization of Sparse Frequency Diverse Array With Time-Invariant Spatial-Focusing Beampattern'"", IEEE Antennas Wireless Propag. Lett., vol.17, no. 12, pp. 2522, Dec. 2018

[18] P. Antonik, M. Wicks, H. Griffiths, et al., "Range dependent beamforming using element level waveform diversity," in *Proc. Int. Waveform Diversity & Design Conf.*, Orlando, USA, 2006, pp. 140–144.

[19] P. Antonik, M. Wicks, H. Griffiths, et al., "Frequency diverse array radars," *in IEEE Conf. Radar*, Verona, NY, USA, 2006, pp. 215–217.

[20] C. Balanis, *Antenna Theory: Analysis and Design*, third ed. New York, NY, USA: Wiley, 2005, pp. 27–32.

[21] L. Huang, X. Li, P. Gong, and Z. He, "Frequency diverse array radar for target range-angle estimation," *Int. J. Computation mathematics in Electrical and Electron. Engineering*, vol. 35 Issue: 3, pp. 1257–1270, 2016.

[22] J. Xu, G. Liao, S. Zhu, L. Huang, and H. C. So, "Joint range and angle estimation using MIMO Radar with frequency diverse array," *IEEE Trans. Signal Processing*, vol. 63, no. 13, pp. 3396–3410, Jul. 2015.

[23] W. Wang, "Overview of frequency diverse array in radar and navigation applications," *IET Radar, Sonar & Navigation*, vol. 10, no. 6, pp. 1001–1012, Jul. 2016.

[24] Y. Ding, J. Zhang, and V. Fusco, "Frequency diverse array OFDM transmitter for secure wireless communication," *Electron. Lett.*, vol. 51, no. 17, pp. 1374–1376, Aug. 2015.

[25] Q. Cheng, J. Zhu, T. Xie, J. Luo, and Z. Xu, "Time-invariant angle-range dependent directional modulation based on time-modulated frequency diverse arrays," *IEEE Access*, vol. 5, pp. 26279–26290, 2017.